\documentclass[notitlepage]{revtex4-1}
\usepackage{setspace}
\usepackage[utf8]{inputenc}
\usepackage[dvipsnames]{xcolor}
\usepackage[colorlinks=true,citecolor=RedViolet,urlcolor=BlueGreen,linkcolor=RedViolet]{hyperref}
\usepackage[normalem]{ulem}
\usepackage{url}
\usepackage{mathrsfs}
\usepackage{amsmath}
\usepackage{graphicx,wrapfig,float,slashed,cancel}
\usepackage{amsmath,amssymb,epsfig,xcolor,stmaryrd}
\usepackage{bm}
\usepackage{enumitem}
\usepackage{hhline,multirow,tabularx} 
\usepackage{graphics}
\usepackage{latexsym}
\usepackage{rotating}
\usepackage{subfigure}
\usepackage{color}
\usepackage{blindtext}
\usepackage{lipsum}
\usepackage{multirow}
\usepackage{physics}
\usepackage{orcidlink}

	\allowdisplaybreaks
\begin{document}
	
	
	\title{\textcolor{BlueViolet}{Impact of QCD sum rules coupling constants on neutron stars structure}}

	\author{F.~Moradi Jangal$^{a}$\orcidlink{0009-0009-9049-0528}}
\email{F.Moradi71@ut.ac.ir }

\author{H.~R.~Moshfegh$^{a,b}$\orcidlink{0000-0002-9657-7116}}
\email{hmoshfegh@ut.ac.ir }
\thanks{Corresponding author}

\author{K.~Azizi$^{a,c}$\orcidlink{0000-0003-3741-2167}}
\email{kazem.azizi@ut.ac.ir}

\affiliation{
	$^{a}$Department of Physics, University of Tehran, North Karegar Avenue, Tehran 14395-547, Iran\\		
	$^{b}$	Centro Brasileiro de Pesquisas Fısicas, Rua Dr. Xavier Sigaud,
	150, URCA, Rio de Janeiro CEP 22290-180, RJ, Brazil	\\	
	$^{c}$Department of Physics, Do\u gu\c s University, Dudullu-\"Umraniye, 34775 Istanbul, T\"urkiye
	}

\date{\today}

\preprint{}
	\begin{abstract}
		We present a detailed investigation on the structure of neutron stars, incorporating the presence of hyperons within a relativistic model under the mean-field approximation. Employing coupling constants derived from QCD sum rules, we explore the particle fraction in beta equilibrium and establish the mass-radius relationship for neutron stars with hyperonic matter. Additionally, we compute the stellar Love number ($\mathcal{K}_{2}$) and the tidal deformability parameter ($\varLambda$), providing valuable insights into the dynamical properties of these celestial objects. Through comparison with theoretical predictions and observational data, our results exhibit good agreement, affirming the validity of our approach. These findings contribute significantly to refining the understanding of neutron star physics, particularly in environments containing hyperons, and offer essential constraints on the equation of state governing such extreme astrophysical conditions.

	\end{abstract}
	
	
	\maketitle
	
	\renewcommand{\thefootnote}{\#\arabic{footnote}}
	\setcounter{footnote}{0}
	
	\section {Introduction}\label{sec:one}
  
 The study of matter under extreme conditions, characterized by high densities and temperatures, has been a major focus in theoretical physics in recent decades. Models and concepts emerging from the study of nuclear matter have significantly contributed to unraveling some mysteries about the strong and weak interaction of matter. Compact stars serve as natural laboratories for investigating such exotic matter, allowing for the examination of processes involving all four known fundamental forces of nature. Existence of Neutron Stars (NSs) was  proposed by Baade and Zwicky in 1934 \cite{Baade:1934wuu}, sparking subsequent development of various models to describe their physics and properties \cite{Chin:1974sa, Huber:1997mg, Heiselberg:1999mq, Lattimer:2000nx, Weber:2004kj}. Theoretical studies are currently being performed more than ever to elucidate neutron star physics through the equation of state (EoS) for dense matter. Despite the numerous assumptions and considerable efforts made to calculate the mass limits and explore the internal composition of neutron stars, several unresolved issues persist, partly due to the likely complex structures of neutron stars. The masses and radii of neutron stars are particular interest in observational experiments. The neutron star population's members with the largest masses and their radii are particularly important for testing neutron star EoSs \cite{Ozel:2016oaf}.
 Due to precise observations of neutron stars, such as the Shapiro delay measurement of a binary millisecond PSR
 J1614-2230 \cite{Demorest:2010bx}, the radius measurement of PSR J0740+6620 \cite{Fonseca:2021wxt} from the Neutron Star Interior Composition Explorer (NICER) and X-ray Multi-Mirror (XMMNewton) data,
 the discovery of PSR J0952-0607 \cite{Romani:2022jhd} by the Low-Frequency Array (LOFAR) radio telescope, and observational data of PSR J0537-6910 \cite{Ho:2015vza,Ho:2020vxt} with NICER timing, we have gained valuable insights into the properties of these celestial objects.
 Additionally, the direct detection of gravitational wave (GW) signals from a binary neutron star merger such as GW170817 \cite{LIGOScientific:2018hze}, and GW190425 \cite{LIGOScientific:2020aai}, observed by Advanced LIGO and  Virgo detectors, have imposed a strong constraint on the mass-radius relation of neutron stars. Especially, the tidal deformability
 of a neutron star \cite{Flanagan:2007ix,Hinderer:2007mb}
 plays an important role in constructing the EoS for neutron star matter. Assuming both NSs in the merger possess have the same EoS, this has led to stringent constraints on the tidal deformability, represented by $\varLambda_{1.4}~=~190_{-120}^{+390} $ and on the radius of a 1.4 solar mass  neutron star, bounded by $11.82 < R_{1.4} < 13.72$ $Km$ NS for GW170817 event \cite{LIGOScientific:2018cki, Lim:2019som,Malik:2018zcf}. Collectively, these groundbreaking detections have provided a substantial corpus of reliable observational data crucial for the investigation of the structure and internal composition of neutron stars \cite{Raaijmakers:2019qny, Miller:2021qha, Chatziioannou:2020pqz, Li:2021crp}. However, a pressing question continues to challenge researchers: the uncertainty regarding whether neutron stars contain additional degrees of freedom, such as hyperons, alongside nucleons. Moreover, the necessary interactions for the emergence of these new degrees of freedom in the context of massive neutron stars remain elusive \cite{Xu:2023gmc}. Consequently, the true internal composition of neutron stars remains an open and compelling question, driving ongoing research efforts in the field. Hyperons may appear in the inner core of neutron stars where densities are around 2 - 3 times the nuclear saturation density, $\rho_{0}$ \cite{Schulze:1998jf,Djapo:2008au,Vidana:2015rsa,Lonardoni:2013gta}. This leads to a softening of the equation of state (EoS) and consequently, the maximum mass of the neutron star could be lower than what current observations suggest, which is about 2~ $M_{\odot}$ \cite{Demorest:2010bx, NANOGrav:2017wvv,NANOGrav:2019jur}. The potential presence of hyperons in the interior of neutron stars poses a significant challenge for our current understanding because it contradicts our observations of high-mass neutron stars. This challenge, often termed the "hyperon puzzle" in the literature, is currently a hot topic of research in nuclear astrophysics.
 
 In this research, we will utilize by using the $\sigma$-$\omega$-$\rho$ model \cite{Chin:1974sa,Mueller:1996pm, Walecka:1974qa,Uechi:2006pz} to describe the EOS of the neutron star matter. The main goal is to explore the impact of coupling constants obtained from the QCD sum rules method on the equation of state of static and non-rotating NSs composed of baryons and leptons. Our aim is to identify a relativistic mean-field model that incorporates various baryon-meson coupling constants to accurately describe nuclear matter and address the hyperon puzzle challenge. The manuscript is organized as follows:
 In Section \ref{sec:two}, we introduce our theoretical model and provide a brief overview of QCD sum rules. Section \ref{sec:three} is dedicated to presenting our results and discussions. Within this section and its subsections, we will compare the predictions from our proposed models with observational data of neutron stars to identify the most suitable set of coupling constants. This comparative analysis will enable us to refine our understanding of the equation of state and gain insights into the internal composition of neutron stars. Finally, we conclude with remarks in Section \ref{sec:four}. 
 
\section {Theoretical Framework}
\label{sec:two}
	\subsection{Lagrangian Density}
This study will utilize a Lagrangian to describe the behavior of particles in the core. The Lagrangian density incorporates terms for leptons, baryons, and mesons, along the interaction terms. It is expressed as follows \cite{Glendenning:1984jr}:
	\begin{equation}
		\mathscr{L} = \sum_{B} \mathscr{L}_{B} + \sum_{\mathcal{M}} \mathscr{L}_{\mathcal{M}} +\sum_{\lambda} \mathscr{L}_{\lambda} + \mathscr{L}_{int}.
	\end{equation}
	Baryons are represented by B, mesons by $\mathcal{M}$, and leptons by $\lambda$.
	The Lagrangian for the core of a NS can be constructed
	using a theory that incorporates three types of mesons: the sigma ($\sigma$) , omega ($\omega$) , and rho ($\rho$). The generalized Lagrangian density, based on the $ \sigma,\omega, \rho $ theory, is:
	\begin{equation}
		\label{eqn:L2}
		\begin{split}
			\mathscr{L}  = & \sum_{B} \bar{\psi}_{B} (i \gamma_{\mu} \partial^{\mu} - m_{B}
			+ g_{BB\sigma}\sigma - g_{BB\omega}\gamma_{\mu}\omega^{\mu} - \dfrac{1}{2}  g_{BB\rho}\gamma_{\mu} 	\boldsymbol{\tau}.\boldsymbol{\rho}^{\mu}) \psi_{B} \\
			& + \dfrac{1}{2}(\partial_{\mu}\sigma\partial^{\mu}\sigma - m^{2}_{\sigma}\sigma^{2}) - \dfrac{1}{4} \omega_{\mu\nu} \omega^{\mu\nu} + \dfrac{1}{2} m^{2}_{\omega} \omega_{\mu} \omega^{\mu}\\
			& -\dfrac{1}{4} \boldsymbol{\rho}_{\mu\nu} . \boldsymbol{\rho}^{\mu\nu} + \dfrac{1}{2} m^{2}_{\rho} \boldsymbol{\rho}_{\mu}.\boldsymbol{\rho}^{\mu} - \dfrac{1}{3} bm_{n} (g_{NN\sigma}\sigma)^{3} - \dfrac{1}{4} c (g_{NN\sigma}\sigma)^{4} \\
			& + \sum_{\lambda} \bar{\psi}_{\lambda}(i \gamma_{\mu} \partial^{\mu} - m_{\lambda})\psi_{\lambda}.\\
		\end{split}
	\end{equation}
	Here, we assume that baryons consist of neutrons, protons, and Hyperons ($\Sigma^{-}$, and $\Lambda$), while mesons include $\sigma, \omega$, and $\rho$. Additionally, leptons are electron and muons. In the Lagrangian density $\partial^{\mu}$ represents the four-derivatives and $\gamma_{\mu}$ are the (covariant) Dirac Matrices. $\boldsymbol{\rho}_{\mu\nu}$ and $\omega_{\mu\nu}$ are the $\rho$ and $\omega$ field strength tensors, respectively. The spinors are shown by $\psi$, while $\bar{\psi}\equiv \psi^{\dagger}\gamma_{0} $ represent their adjoint. The coupling constants in the above Lagrangian density are shown with $g_{BB\sigma}$, $g_{BB\omega}$, $g_{BB\rho}$, b, and c, that indicate the interactions between baryons and mesons as well as scalar self-interactions (b and c). In addition, $\boldsymbol{\tau}$ are the isospin matrices.
	A relativistic mean-field approximation (MFA) is employed to investigate high density nuclear and neutron matter in this study. To parameterise the density dependence of the energy functional, nonlinear interactions between the fields are introduced based on idea from effective field theory \cite{Mueller:1996pm}. Specially, we focus on different types of nonlinearity involving scalar-isoscalar ($\sigma$), vector-isoscalar ($\omega$), and vector-isovector ($\rho$) fields. Within the MFA framework, it is assumed that the particles do not interact with each other but rather experience an average effect from surrounding of the system.
	\subsection{Chemical Potential and Meson Fields}
	As noted by Glendenning \cite{Glendenning:1984jr}, the chemical equilibrium problem can be addressed without considering individual reactions. For frozen static matter, the chemical potential ($\mu$) represents the energy of a particle at the highest level within the Fermi sea. By solving the Euler-Lagrange equations and using mean-field approximation for the meson field, we can determine the fields and their eigenvalues. This allows us to express the baryon fields in momentum representation as:
	\begin{equation}
		[\gamma_{\mu}(k^{\mu}-g_{BB\omega}\omega^{\mu}-\dfrac{1}{2}g_{BB\rho}\boldsymbol{\tau}.\boldsymbol{\rho}^{\mu})-(m_{B}-g_{BB\sigma}\sigma)]\psi_{B}(k)=0,
	\end{equation}
    and the eigenvalues for particles and antiparticles given by:
	\begin{equation}
		\label{Eq:ch2}
		e_{B}(k) = g_{BB\omega}\omega_{0} + g_{BB\rho}\rho_{03}I_{B3} + \sqrt{k^{2}+(m_{B}-g_{BB\sigma}\sigma)^{2}},
	\end{equation}
	\begin{equation}
		\bar{e}_{B}(k) = -g_{BB\omega}\omega_{0} - g_{BB\rho}\rho_{03}\bar{I}_{B3} + \sqrt{k^{2}+(m_{B}-g_{BB\sigma}\sigma)^{2}}.
	\end{equation}
	Where, $I_{B3}$ represents the isospin projection for baryon B. In addition, $\sigma, \omega_{0}$ and $\rho_{03}$ are meson fields in the uniform static matter, and we can write them as:
	\begin{equation}
		\label{eqn:m-f-one}
		\omega_{0} = \sum_{B} \dfrac{g_{BB\omega}}{m_{\omega}^{2}}\rho_{B},
	\end{equation}
	\begin{equation}
		\label{eqn:m-f-two}
		\rho_{03} = \sum_{B} \dfrac{g_{BB\rho}}{m_{\rho}^{2}}I_{B3}\rho_{B},
	\end{equation}
	\begin{equation}
		\label{eqn:m-f-tree}
		m_{\sigma}^{2}\sigma = - b m_{N}g_{NN\sigma}(g_{NN\sigma}\sigma)^{2} - c g_{NN\sigma}(g_{NN\sigma}\sigma)^{3}\\
		+\sum_{B}\dfrac{2J_{B}+1}{2\pi^{2}} g_{BB\sigma}\int_0^{k_{B}} \dfrac{m_{B}-g_{BB\sigma}\sigma}{\sqrt{k^{2}+(m_{B}-g_{BB\sigma}\sigma)^{2}}}k^{2}dk.
	\end{equation}
	
	The spin of a baryon is denoted by $J_{B}$. The total baryon number density ($\rho_{B}$) is related to the Fermi momentum ($k_{F}$) by the following relation,
	\begin{equation}
		\rho_{total} = \sum_{i} \rho_{i} = \sum_{i} \dfrac{k_{F_{i}}^{3}}{3\pi^{2}},
	\end{equation}
	where, $i$ is the set of baryons in the model.
	\subsection{Equation of State and Equilibrium Conditions}
 By applying the Energy-Momentum tensor to the Lagrangian density (Eq. \ref{eqn:L2}), we obtain the following expressions for energy and pressure per particle:
	\begin{equation}
		\begin{split}
			\varepsilon =& \dfrac{1}{3} b ~ m_{N}(g_{NN\sigma}\sigma)^{3} +\dfrac{1}{4} c (g_{NN\sigma}\sigma)^{4}+ \dfrac{1}{2}m_{\sigma}^{2}\sigma^{2} + \dfrac{1}{2}m_{\omega}^{2}\omega_{0}^{2} + \dfrac{1}{2}m_{\rho}^{2}\rho_{03}^{2} \\
			&+\sum_{B}\dfrac{2J_{B}+1}{2\pi^{2}}\int_0^{k_{B}} \sqrt{k^{2}+(m_{B}-g_{BB\sigma}\sigma)^{2}}k^{2}dk \\
			&+\sum_{\lambda}\dfrac{1}{\pi^{2}}\int_0^{k_{\lambda}}\sqrt{k^{2}+m_{\lambda}^{2}}k^{2}dk ,\\
		\end{split}
	\end{equation}
	\begin{equation}
		\begin{split}
			p =& -\dfrac{1}{3} b ~ m_{N}(g_{NN\sigma}\sigma)^{3} -\dfrac{1}{4} c (g_{NN\sigma}\sigma)^{4}- \dfrac{1}{2}m_{\sigma}^{2}\sigma^{2} + \dfrac{1}{2}m_{\omega}^{2}\omega_{0}^{2} + \dfrac{1}{2}m_{\rho}^{2}\rho_{03}^{2} \\
			&+\dfrac{1}{3}\sum_{B}\dfrac{2J_{B}+1}{2\pi^{2}}\int_0^{k_{B}}\dfrac{k^{4}dk}{ \sqrt{k^{2}+(m_{B}-g_{BB\sigma}\sigma)^{2}}}\\
			&+\dfrac{1}{3}\sum_{\lambda}\dfrac{1}{\pi^{2}}\int_0^{k_{\lambda}}\dfrac{k^{4}dk}{\sqrt{k^{2}+m_{\lambda}^{2}}}.\\
		\end{split}
	\end{equation}
	where, $ m_{N} $ and $ m_{B}$ are mass of nucleon and baryons, respectively. The masses of the mesons are denoted by $m_{\sigma}$, $m_{\omega}$, and $m_{\rho}$. This study employs a model of a cold NS comprising a combination of neutrons, protons, electrons, muons, and hyperons ($\Sigma^{-}$ and $\Lambda$). Due to the high rest mass of the $\tau$ lepton, its contribution is considered negligible. The possible processes to establish the beta equilibrium condition in the system are as follows:
	\begin{equation}
		\label{U1}
		n \longrightarrow p + e + \bar{\nu}_{e},
	\end{equation}
	and
	\begin{equation}
		\label{U2}
		p + e \longrightarrow n + \nu_{e}.
	\end{equation}
	We know these processes as "Urca processes" \cite{Haensel1995}. When the energy of electrons increases enough, muons can be produced as,
	\begin{equation}
		e \longrightarrow \mu + \bar{\nu}_{\mu} + \nu_{e}.
	\end{equation}
	The appearance of each hyperon species in matter depends on the electric charge and isospin. Since nuclear matter has an excess of positive charge and negative isospin, the formation of hyperons is favoured by negative charge and positive isospin, as well as lower mass. The appearance of each hyperon species at a specific baryon number density is typically the outcome of various factors interacting with each other. However, a detailed quantitative analysis necessitates modeling the interactions at high densities \cite{Balberg:1998ug}.
	At higher densities, $\Sigma^{-}$ and $\Lambda$ could be produced via the following processes,
	\begin{equation}
		\label{eq:Sig}
		n + e \longrightarrow \Sigma^{-} + \nu_{e},
	\end{equation}
	and
	\begin{equation}
		\label{eq:Lamb}
		n + n \longrightarrow n + \Lambda.
	\end{equation}
	The thermodynamic equilibrium conditions for chemical potentials that govern the above equations are as follows:
	\begin{equation}
		\label{eqn:mu}
		\mu_{i} = B_{i} \mu_{n} - Q_{i} \mu_{e}.
	\end{equation}	
Where, $B_{i}$, and $Q_{i}$ denote the baryon charge and electric charge of each species. For Eqs. (\ref{U1}) and (\ref{U2}), the chemical potential of neutrinos is zero since they escape the star as it evolves. 
Additionally, the system maintains charge neutrality, which is established internally, as:
	\begin{equation}
		\label{eqn:ch-n}
		\rho_{p} = \rho_{e} + \rho_{\mu} + \rho_{\Sigma^{-}},
	\end{equation}
	
where $\rho_{i}$ is the number density of each particle. Finally, number density conservation is expressed as:
	\begin{equation}
		\label{eqn:b-d}
		\rho_{B} = \rho_{n} + \rho_{p} + \rho_{\Sigma^{-}} +  \rho_{\Lambda}.
	\end{equation}
	
	\subsection{Coupling Constants}
    We have incorporated various baryon-meson coupling constants in the equation of state to describe stellar matter. Two types of coupling constants have been utilized. One type of parameters was extracted from Refs. \cite{Glendenning:1991es, Karimi:2022vxw}, which is based on the $\sigma$-$\omega$-$\rho$ model. The other method is based on the QCD sum rules, providing a more fundamental based approach.
	Quantum Chromodynamics Sum Rules (QCDSR), is a method that connects the hadronic and the QCD descriptions of matter. It was introduced in Ref. \cite{Shifman:1978bx} for mesons and in Ref. \cite{Ioffe:1981kw} for baryons, and has since become a powerful tool for the phenomenology of hadronic physics. Various hadronic properties  can be determined via QCDSR, such as masses, decay rates, coupling constants, magnetic moments and weak form factors.  These properties are expressed in terms of QCD parameters, such as quark masses, and quark and gluon condensates \cite{Aliev:2009jt,Agaev:2016dev,Azizi:2016dhy}.
	The main idea of QCDSR is to construct a correlation function of two/three hadronic currents, which can be calculated in two different ways: on one hand, by using the operator product expansion (OPE) and the QCD perturbation theory at short distances; and on the other hand, by using the hadronic spectral representation and the dispersion relation at long distances \cite{Doi:2003cd}.
	The idea behind the OPE method is that the properties of a hadron, such as its mass or decay constant, can be related to the behavior of quarks and gluons inside the hadron through a series expansion. This expansion expresses a physical quantity, such as the mass of a hadron, as a sum of terms involving quark and gluon field operators of increasing dimension, with coefficients known as Wilson coefficients. 
	As mentioned, one of the applications of QCDSR method is to calculate the meson-baryon coupling constants, which are essential for understanding the hadronic interactions. For instance, to determine the coupling constant of a vector meson to the nucleon, one can consider the following three-point correlation function:
	\begin{equation}
		\Pi^{\mu}(q) = i^{2} \int d^4x \int d^4y ~ e^{-ip\cdot x} ~ e^{ip^{\prime}\cdot y} ~ \langle 0 | \mathcal{T}\big(J_N(y) ~ J^{\mu}_{\mathcal{M}}(0) ~ \bar{J}_N(x)\big) | 0 \rangle,
	\end{equation}
	where, $\mathcal{T}$ is the time ordering operator and $q = p - p^{\prime}$ is the transferred momentum. $J_N$ and $J^{\mu}_{\mathcal{M}}$ denote the interpolating fields of the nucleon and meson, respectively. Their interpolating currents in terms of the quark fields, are:
	\begin{equation}
		J^{\mu}_{\mathcal{M}}(0) = \bar{q}_{1}(0) ~\gamma^{\mu}~ q_{2}(0),
	\end{equation}
	and
	\begin{equation}
		J_N(y) = \epsilon_{\alpha \beta z} \big( u^{\alpha^{T}}(y)~ C~ \gamma_\nu ~ u^{\beta}(y) \big) \gamma_5~ \gamma^\nu ~d^{z}(y).
	\end{equation}
	Here, $q_{i}$ are the light quarks, and $\alpha$, $\beta$, $z$ are the color indices, $C$ is the charge conjugation operator, and $T$ denotes the transpose with respect to the Dirac indices. By applying the OPE and the spectral representation on this correlation function, one can obtain a sum rule for the nucleon-nucleon-meson coupling constant. The Borel transformation is applied to both sides to suppress the contributions of the higher states and continuum \cite{Azizi:2015bxa}.
	The above procedure can be applied to calculate the baryon-meson couplings, such as the $NN\omega$, $NN\rho$, $ \Lambda \Lambda \rho$, $\Lambda\Lambda\omega$, etc. The results can be compared with the empirical values obtained from the One-Boson Exchange (OBE) models of the two-baryon interactions, as well as the SU(3)-flavor relations and the chiral symmetry constraints \cite{Aliev:2009ei,Wang:2007yt,Erkol:2006sa,Erkol:2006eq,Zamiralov:2013gva}.
	The accuracy of the coupling constant calculation using QCDSR depends on a number of factors, such as  the quality of the assumptions made in the OPE method and working intervals of some auxiliary parameters entering the sum rules. However, QCDSR have been successfully employed to calculate coupling constants for a broad range of processes involving hadrons, and have offered important insights into the behavior of quarks and gluons within QCD.
	In this study, we use the coupling constants obtained by two different methods ( $\sigma-\omega-\rho$ Model and QCDSR method) as described in Refs. \cite{Karimi:2022vxw,Aliev:2009ei,Wang:2007yt,Erkol:2006sa,Erkol:2006eq,Zamiralov:2013gva}, which are listed in Table \ref{table:cc2}, and divided into four sets according to Table \ref{table:D-G}.

	\begin{table}[H]
		\caption{Meson-baryon coupling constants (dimensionless).}
		\label{table:cc2}
		\small
		\begin{center}
			\begin{tabular}{|| c ||  c | c | c | c | c | c ||}
				\hline
				
				& $\sigma$-$\omega$-$\rho$ base &\multicolumn{5}{|c||}{QCDSR method} \\
				\cline{2-7}
				
				C.C & Ref. \cite{Karimi:2022vxw} & Ref. \cite{Aliev:2009ei}  & Ref. \cite{Wang:2007yt} &Ref. \cite{Erkol:2006sa} &Ref. \cite{Erkol:2006eq} & Ref. \cite{Zamiralov:2013gva}  \\
				
				& $\zeta=-2$ &  &  & & &  \\ 
				\hline\hline
				$g_{NN\sigma}$  & 8.3$\pm$ 0.0 & - & - & - & 14.4 $\pm$ 3.7 & -  \\ 
				$g_{NN\omega}$  & 8.6$\pm$ 0.0 & -8.9$\pm$ 1.1 & - & 7.2$\pm$ 1.8 & - & 6.5$\pm$ 1.1  \\
				$g_{NN\rho}$   & 8.5$\pm$ 0.0 & -2.5$\pm$ 1.1 & 3.2$\pm$ 0.9 & 2.4$\pm$ 0.6 & - & 3.0$\pm$ 1.0  \\
				$g_{\Lambda\Lambda\sigma}$  & 5.9$\pm$ 0.0 & - & - & - & 7.0$\pm$ 1.9 & - \\  
				$g_{\Lambda\Lambda\omega}$  & 6.1$\pm$ 0.0 & -7.1$\pm$ 1.1 & - & 4.8$\pm$ 1.2 & - & 2.8$\pm$ 0.4 \\
				$g_{\Lambda\Lambda\rho}$  & 6.0$\pm$ 0.0 & - & - & 0.0 & - & -  \\
				$g_{\Sigma^{-}\Sigma^{-}\sigma}$ & 5.1$\pm$ 0.0 & - & - & - & - & -  \\
				$g_{\Sigma^{-}\Sigma^{-}\omega}$ & 5.3$\pm$ 0.0 & - & - & - & - & -   \\
				$g_{\Sigma^{-}\Sigma^{-}\rho}$  &5.2$\pm$ 0.0&  - & - & - & - & -   \\
				$g_{\Sigma^{+}\Sigma^{+}\sigma}$  &5.2$\pm$ 0.0 & - & - & - & 14.1$\pm$ 3.7 & -  \\
				$g_{\Sigma^{+}\Sigma^{+}\omega}$   & 5.3$\pm$ 0.0 & 6.6$\pm$ 1.0 & - & 4.8$\pm$ 1.2 & - & 4.6$\pm$ 1.0  \\
				$g_{\Sigma^{+}\Sigma^{+}\rho}$  & 5.3$\pm$ 0.0 & 7.2$\pm$ 1.2 & 4.0$\pm$ 1.0 & 4.8$\pm$ 1.2 & - & 4.6$\pm$ 1.0  \\ 
				\hline
			\end{tabular}
		\end{center}
	\end{table}
	
	\begin{table}[H]
		\caption{Coupling constant sets.}
		\label{table:D-G}
		\small
		\begin{center}
			\begin{tabular}{|| c || c | c | c | c | c | c | c | c | c | c ||}
				\hline
				Sets & 	$g_{NN\sigma}$ & $g_{NN\omega}$ & 	$g_{NN\rho}$& 	$g_{\Lambda\Lambda\sigma}$ & 	$g_{\Lambda\Lambda\omega}$  & 	$g_{\Lambda\Lambda\rho}$  &	$g_{\Sigma^{-}\Sigma^{-}\sigma}$ & 	$g_{\Sigma^{-}\Sigma^{-}\omega}$ & 	$g_{\Sigma^{-}\Sigma^{-}\rho}$  & Reference \\ 
				\hline\hline
				Set 1 & 8.3$\pm$ 0.0 & 8.6$\pm$ 0.0& 8.5$\pm$ 0.0& 5.9$\pm$ 0.0& 6.1$\pm$ 0.0& 6.0$\pm$ 0.0& 5.1$\pm$ 0.0& 5.3$\pm$ 0.0& 5.2$\pm$ 0.0& Ref. \cite{Karimi:2022vxw}  \\
				Set 2  & 14.4 $\pm$ 3.7 & -8.9$\pm$ 1.1 & -2.5$\pm$ 1.1 &7.0$\pm$ 1.9& -7.1$\pm$ 1.1& 0.0 & 14.1$\pm$ 3.7& 6.6$\pm$ 1.0& 7.2$\pm$ 1.2& Refs. \cite{Aliev:2009ei, Erkol:2006eq}  \\ 
				
				Set 3 & 14.4 $\pm$ 3.7 & 7.2$\pm$ 1.8 & 2.4$\pm$ 0.6 &7.0$\pm$ 1.9& 4.8$\pm$ 1.2& 0.0 & 14.1$\pm$ 3.7&  4.8$\pm$ 1.2&  4.8$\pm$ 1.2&  Refs. \cite{Erkol:2006sa, Erkol:2006eq}  \\ 
				
				Set 4 & 14.4 $\pm$ 3.7 & 6.5$\pm$ 1.1 & 3.0$\pm$ 1.0 &7.0$\pm$ 1.9& 2.8$\pm$ 0.4& 0.0 & 14.1$\pm$ 3.7&  4.6$\pm$ 1.0&  4.6$\pm$ 1.0& Refs. \cite{Zamiralov:2013gva, Erkol:2006eq}   \\ 
				 
				\hline
			\end{tabular}
		\end{center}
	\end{table}

\section {Results and discussion}\label{sec:three}	
\label{sec:result}
\subsection{Particle Fractions and EOS}

\begin{figure}
	\centering
	\subfigure[]{\label{fig:1}
	\includegraphics*[width=.48\linewidth, height=0.23\paperheight]{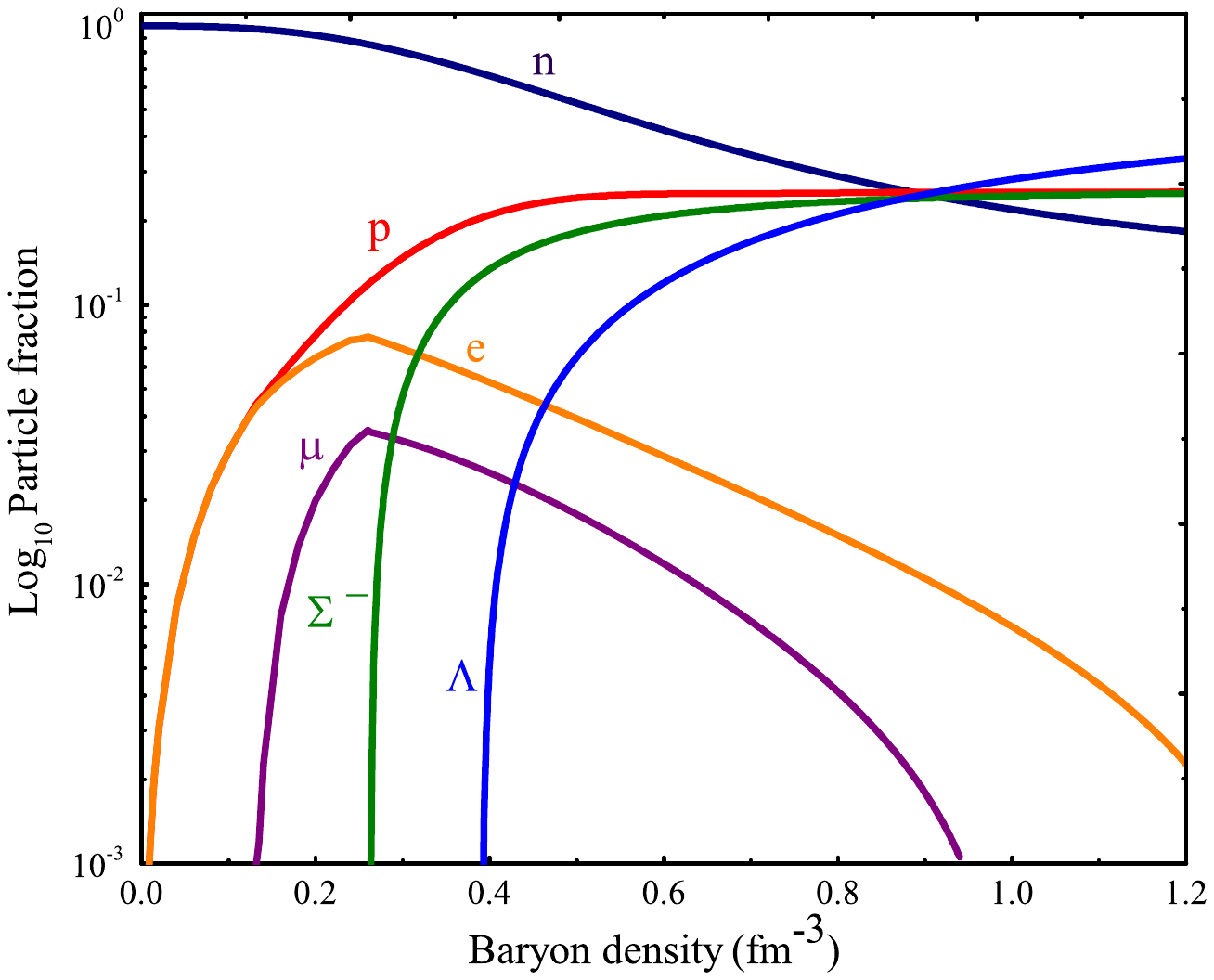}}
	\hspace{1mm}
\subfigure[ ]{\label{fig:2}
	\includegraphics*[width=.48\linewidth]{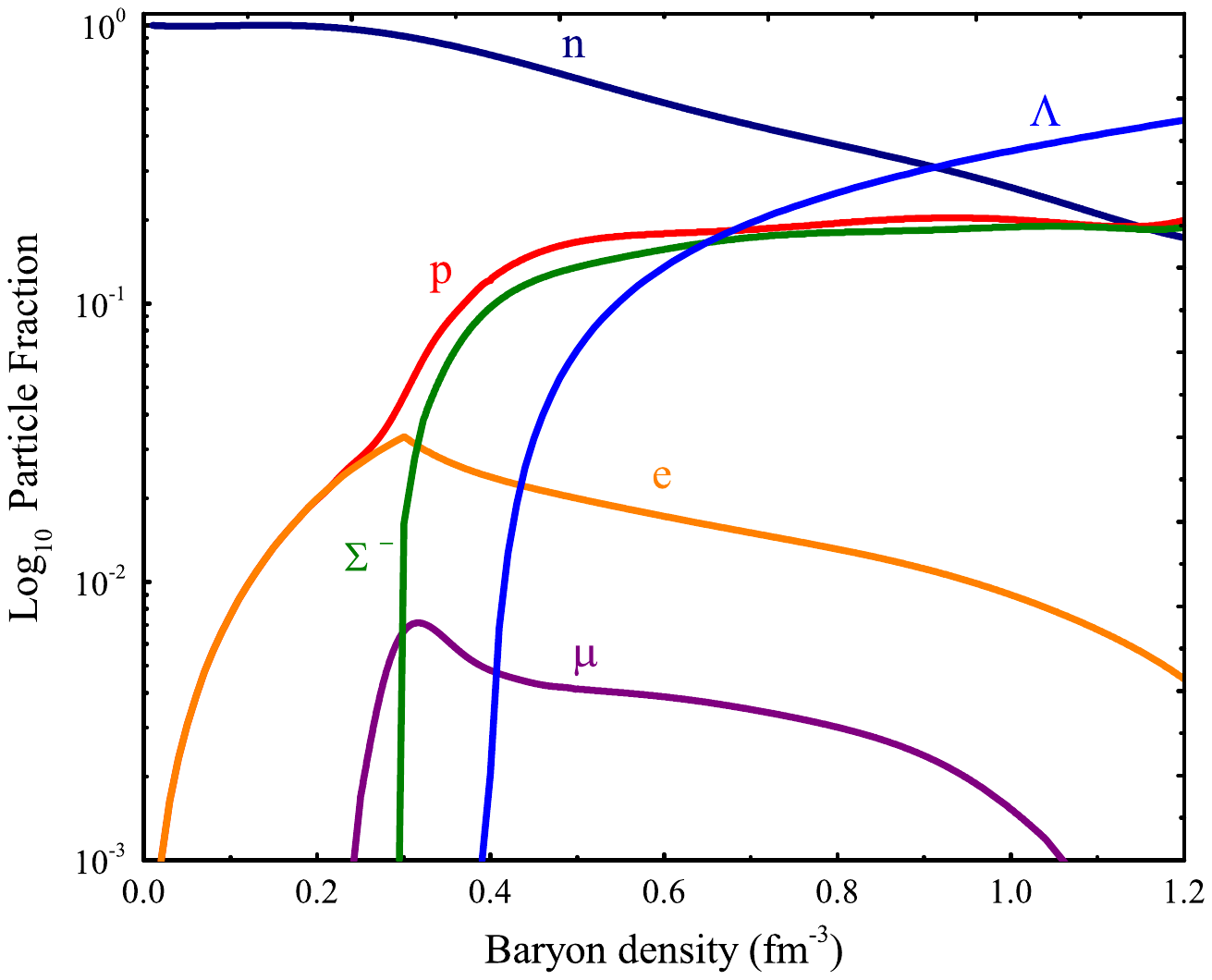}}
	\hspace{1mm}
\subfigure[ ]{\label{fig:3}
	\includegraphics*[width=.48\linewidth]{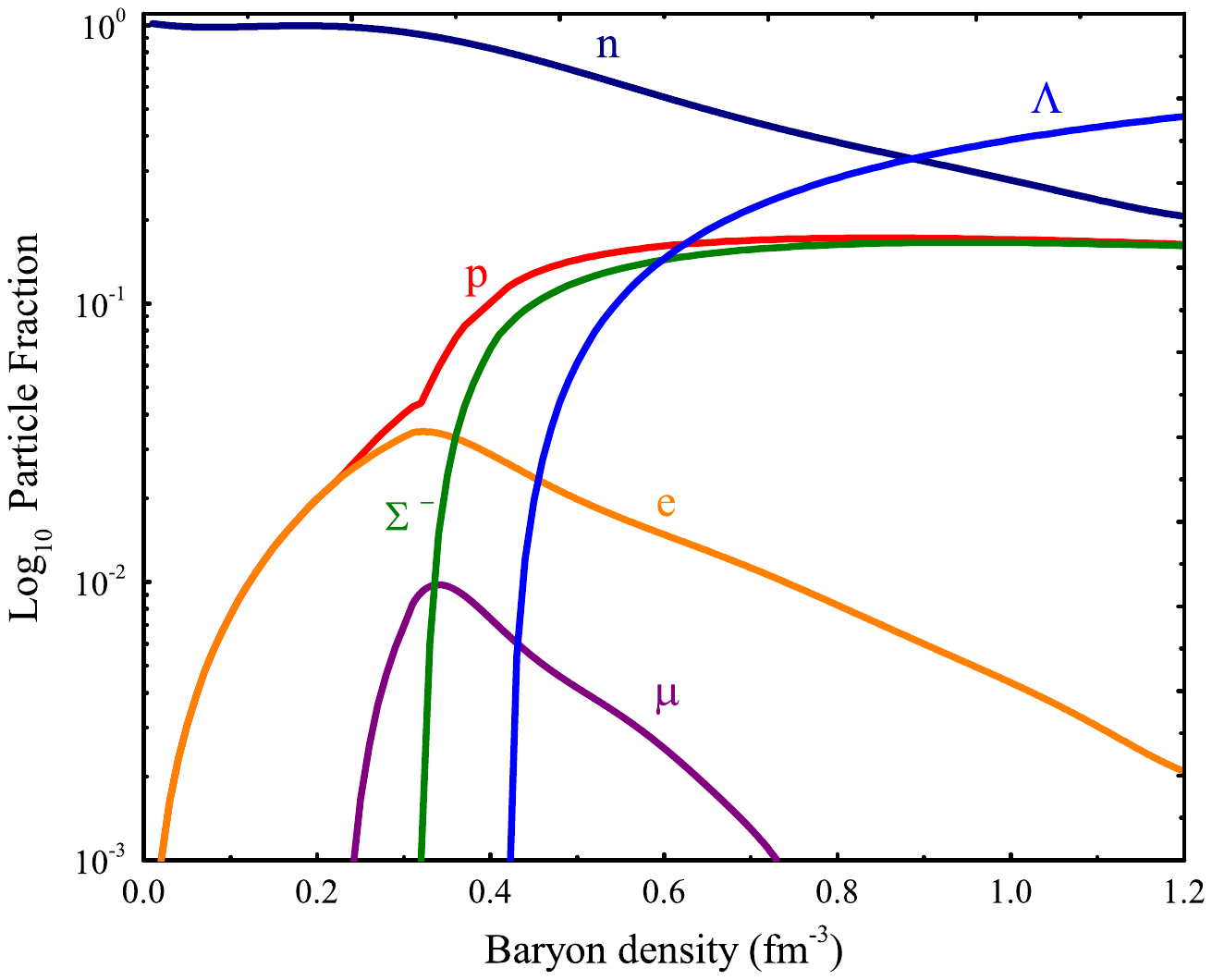}}
		\hspace{1mm}
	\subfigure[ ]{\label{fig:4}
		\includegraphics*[width=0.48\linewidth]{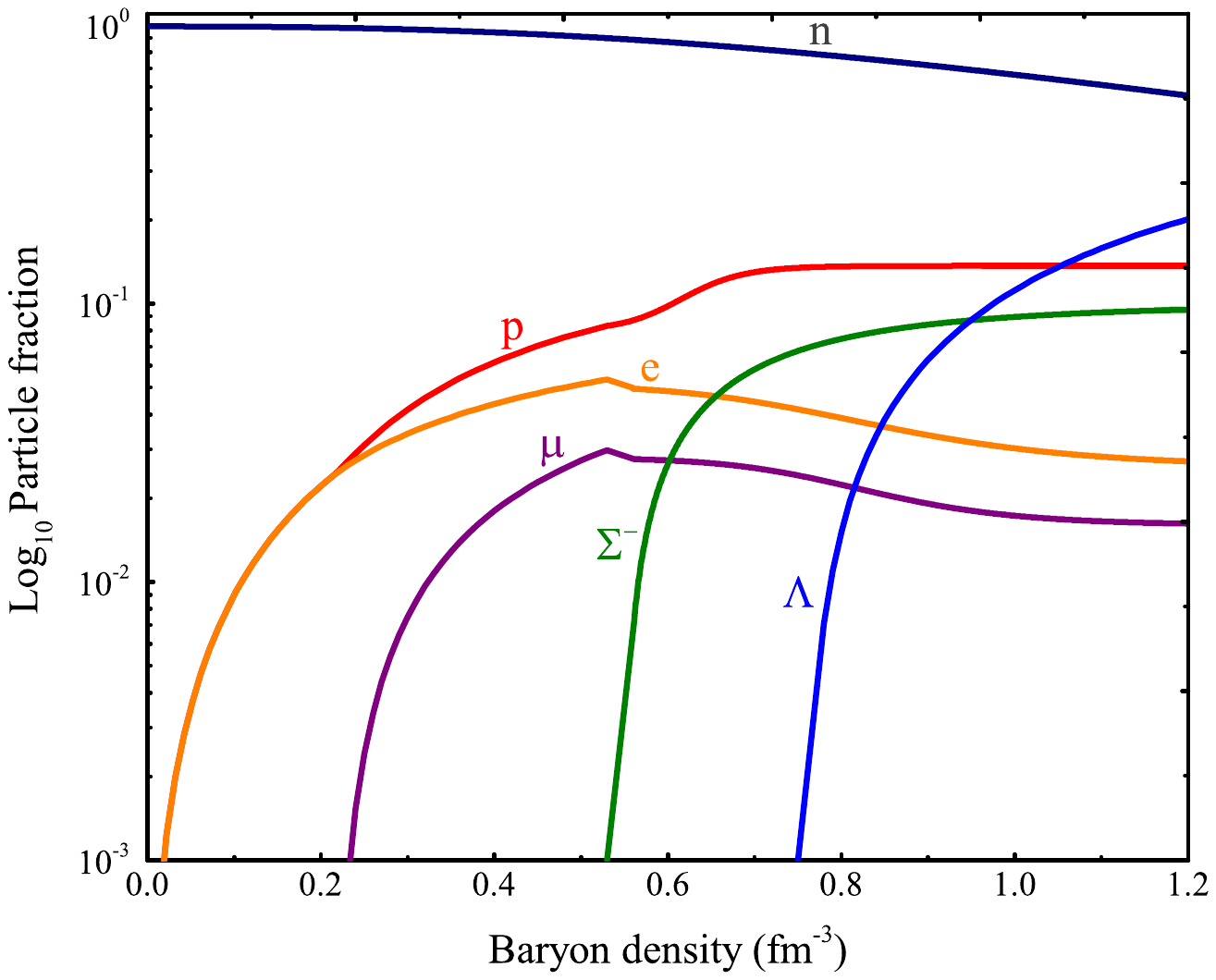}}
	\caption{Particle fraction $Y_{i}$ for npYe$\mu$ system using coupling constants of (a): set~1, (b): set~2, (c): set~3, and (d): set~4.}
	\label{4fig}
\end{figure}

\begin{figure}[H]
	\centering
	\subfigure[]{\label{fig:p-e-tot}
		\includegraphics*[width=.48\linewidth, height=0.23\paperheight]{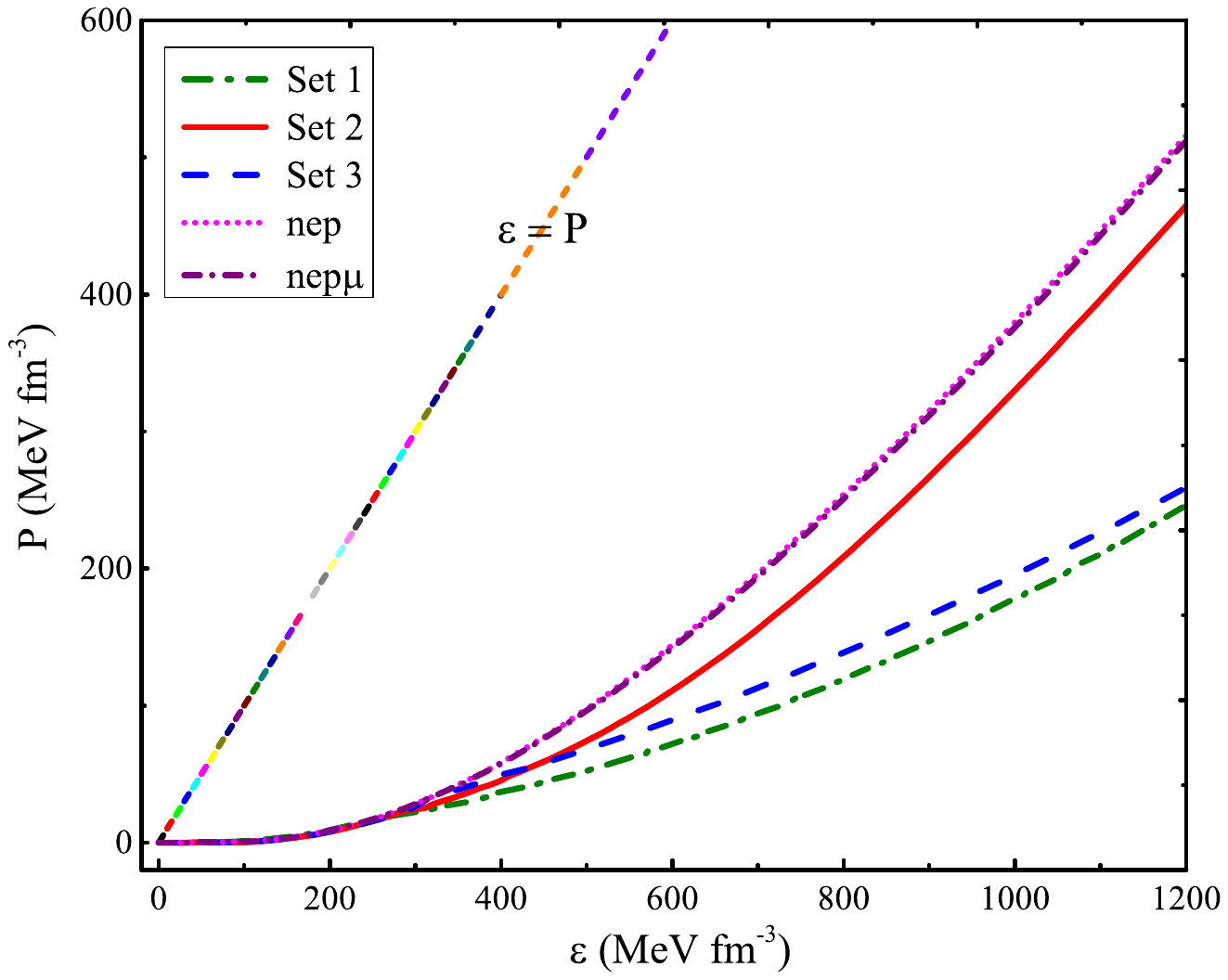}}
	\hspace{1mm}
	\subfigure[]{\label{fig:p-n-tot}
		\includegraphics*[width=0.48\linewidth]{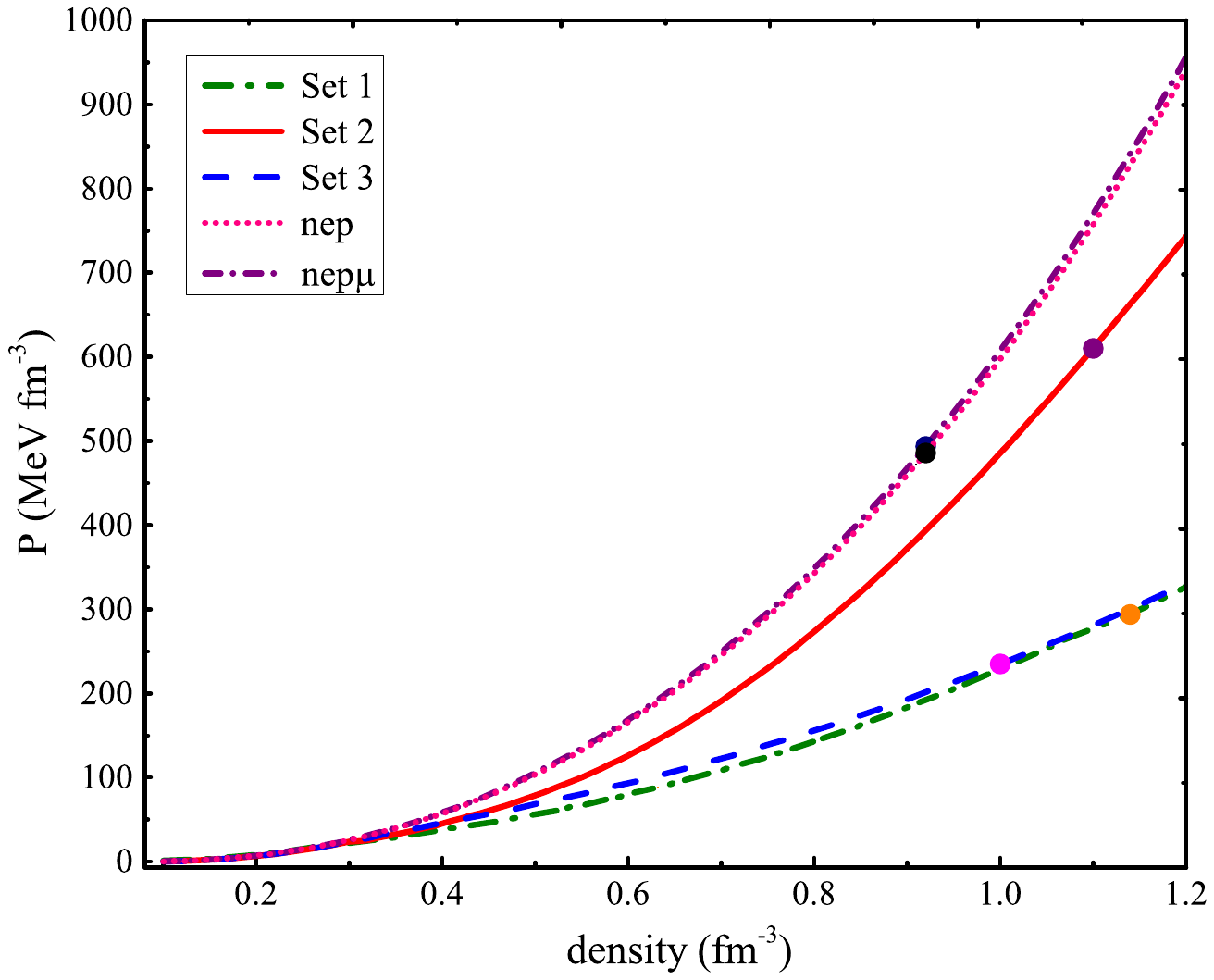}}
	\hfill \\
	\caption{(a): Pressure vs. energy density, (b): Pressure vs. density for npYe$\mu$ systems using the various sets of coupling constants.}
	{\label{2fig}}
\end{figure}

  To explore the structure and composition of a neutron star, it's essential to determine the number density of each potential particle within the star. Thus, we assess the number density of each particle relative to the baryonic number density, representing it as a fraction of particles. To obtain the particle fractions in the NS matter, we solve the sets of equations of chemical potentials ($\ref{eqn:mu}$), charge neutrality ($\ref{eqn:ch-n}$), baryon number density ($\ref{eqn:b-d}$) and meson fields $(\ref{eqn:m-f-one})-(\ref{eqn:m-f-tree})$ at varying densities.  The relative particle fractions as a function of the baryon number density of each coupling constant sets are presented in Fig. $\ref{4fig}$. This figure illustrates that when the electron chemical potential reaches a threshold equal to or surpassing the mass of muon, muon can emerge within the system. As depicted in the figure, the threshold for the onset of muon within the set of coupling constants derived from the QCDSR method (sets 2-4) is approximately around a baryon number density of 0.23 $fm^{-3}$, exceeding what was obtained in \cite{Karimi:2022vxw} using a simple parameterization of coupling constants. At low densities, the system is chiefly composed of neutron matter, but at high densities, there is a significant contribution from other baryons.
  The emergence of hyperons in NS cores transpires as the nucleon chemical potentials surpass the mass differences between nucleons and hyperons, while the threshold for hyperon appearance is determined by their interactions. As the Hyperon fraction increase, the nucleon fractions decrease due to the baryon number density conservation (Eq. $\ref{eqn:b-d}$). The particle fractions become particularly intriguing with the presence of hyperons ($\Sigma^{-}$ and $\Lambda$) in the system. With the appearance of the $\Sigma^{-}$ hyperon, the density of negatively charged leptons begins to decrease rapidly. This occurs because the charge neutrality condition (Eq. $\ref{eqn:ch-n}$) is now fulfilled primarily by the $\Sigma^{-}$ hyperon. Similarly, the emergence of the $\Lambda$ hyperon will accelerate the disappearance of neutrons, as both are neutral particles. The $\Sigma^{-}$ fraction reaches a saturation approximately 0.1. However, in the case of set~1 in Fig. $\ref{fig:1}$, this saturation occurs around 0.2. Meanwhile, the $\Lambda$ particles, unaffected by isospin-dependent forces, continue to increase until they are eventually saturated by short-range repulsion forces. The onset of hyperons is met when the sum of the chemical potentials of neutrons and electrons is greater than or equal to the mass of $\Sigma^{-}$ ($\mu_{n}$ +  $\mu_{e}$ $\geq$ $m_{\Sigma^{-}}$), and when the chemical potential of neutrons is greater than or equal to the mass of $\Lambda$, respectively.
  With the exception of the configuration in set~4 (Fig. $\ref{fig:4}$), the muons are entirely annihilated, and the electron fraction drops below 1 $\% $. In contrast, the electron fraction exceeds 10 $\% $ in nucleonic matter. It is widely acknowledged that hyperons start to emerge at a density of approximately 2-3 times the nuclear saturation density ($\rho_{0}$) \cite{Schulze:1998jf,Djapo:2008au,Vidana:2015rsa,Lonardoni:2013gta}, with the threshold values being influenced by the nature of their interactions.
  The threshold values for the presence of muons and hyperons for each set of coupling constants studied in this research are provided in Table \ref{table:threshod}. These values consistent with other studies' reports, such as Refs. \cite{Schulze:1998jf,Djapo:2008au,Vidana:2015rsa,Lonardoni:2013gta}. However, they exceed the usual reported values for set~4 of coupling constants. As seen in Table \ref{table:D-G}, the coupling constants of set~4 for $\omega$ and $\rho$ meson exchanges are lower than those for the other sets. Consequently, set~4 exhibits weaker interactions for hyperons compared to the other sets. This could explain the onset of hyperons at higher baryon number densities in set~4.
  However, such high baryon densities only occur within a small radius around the center of a neutron star. Therefore, the amount of hyperons does not play a significant role in determining the overall structure of the NS. 
  Results clearly demonstrate the impact of coupling constants on the system, because, according to Eqs. (\ref{Eq:ch2}) to (\ref{eqn:m-f-tree}), the values of the coupling constants affect the values of the meson fields which is turn affect the chemical potential. This can lead to different EOS and physics for each system.
\begin{table}[H]
	\caption{The threshold values (in $fm^{-3}$) for the presence of muons and hyperons in each set.}
	\label{table:threshod}
	\small
	\begin{center}
		\begin{tabular}{|| c || c | c | c ||}
			\hline
			 & $\mu$ & $\Sigma^{-}$ & $\Lambda$  \\ 
			\hline\hline
			Set 1 & 0.13 & 0.27 & 0.39  \\ 
			Set 2 & 0.23 & 0.29 & 0.39  \\ 
			Set 3 & 0.23 & 0.32 & 0.42  \\ 
			Set 4 & 0.22 & 0.53 & 0.76  \\ 
			\hline
		\end{tabular}
	\end{center}
\end{table}
The equation of state plays a crucial role in investigating the properties of NSs. Therefore, we present our calculated results with the presentation of our EOS findings. Fig. \ref{fig:p-e-tot} shows the EOS for NSs when $\Sigma^{-}$ and $\Lambda$ hyperons are present for different sets of coupling constants, as well as for NS matter without hyperons. The line labeled "$P = \varepsilon$" corresponds to the causal limit, where the speed of sound is equal to the speed of light. At low densities, the curves coincide, indicating the existence of only nucleonic matter in $\beta$-equilibrium within the systems.
The pressure-density diagram for all configurations is shown in Fig. \ref{fig:p-n-tot}. The bold points on the curves indicate the central densities in the stars with the maximum mass. We have checked the properties of nuclear matter at saturation point too. We observed that all investigated EOS can reproduce the saturation properties like saturation density and energy as well as incompatibility at saturation point in acceptable range. So in this study we focus on EOS predictions at high densities. The pressure-density relation contains all the essential information necessary for determining the macroscopic properties such as mass, radius, stability and gravitational wave emission of a NS. The pressure-density also determines the internal structure of NSs, such as the presence of different phases of matter including hyperons, quarks, or super fluids, in the core or the crust of the star. The stiffness of the equation of state is crucial in determining the maximum mass of a neutron star. A stiffer equation of state typically leads to a higher maximum mass for the star. In the figure, despite the presence of hyperons in the star, the equation of state is softer compared to the scenario where hyperons are absent. However, as depicted in the figure, the equations of state resulting from set 2 of coupling constants is sufficiently stiff to satisfy observational constraints related to maximum mass, surpassing 2 times the mass of the Sun. This observation is significant in addressing the hyperon puzzle, and we will explore this further in the following sections.
\subsection{Mass-Radius relation of Neutron Stars}
	\label{TOV}
\begin{figure}[H]
	\centering
	\subfigure[]{\label{fig:r-m-tot}
		\includegraphics*[width=.48\linewidth, height=0.23\paperheight]{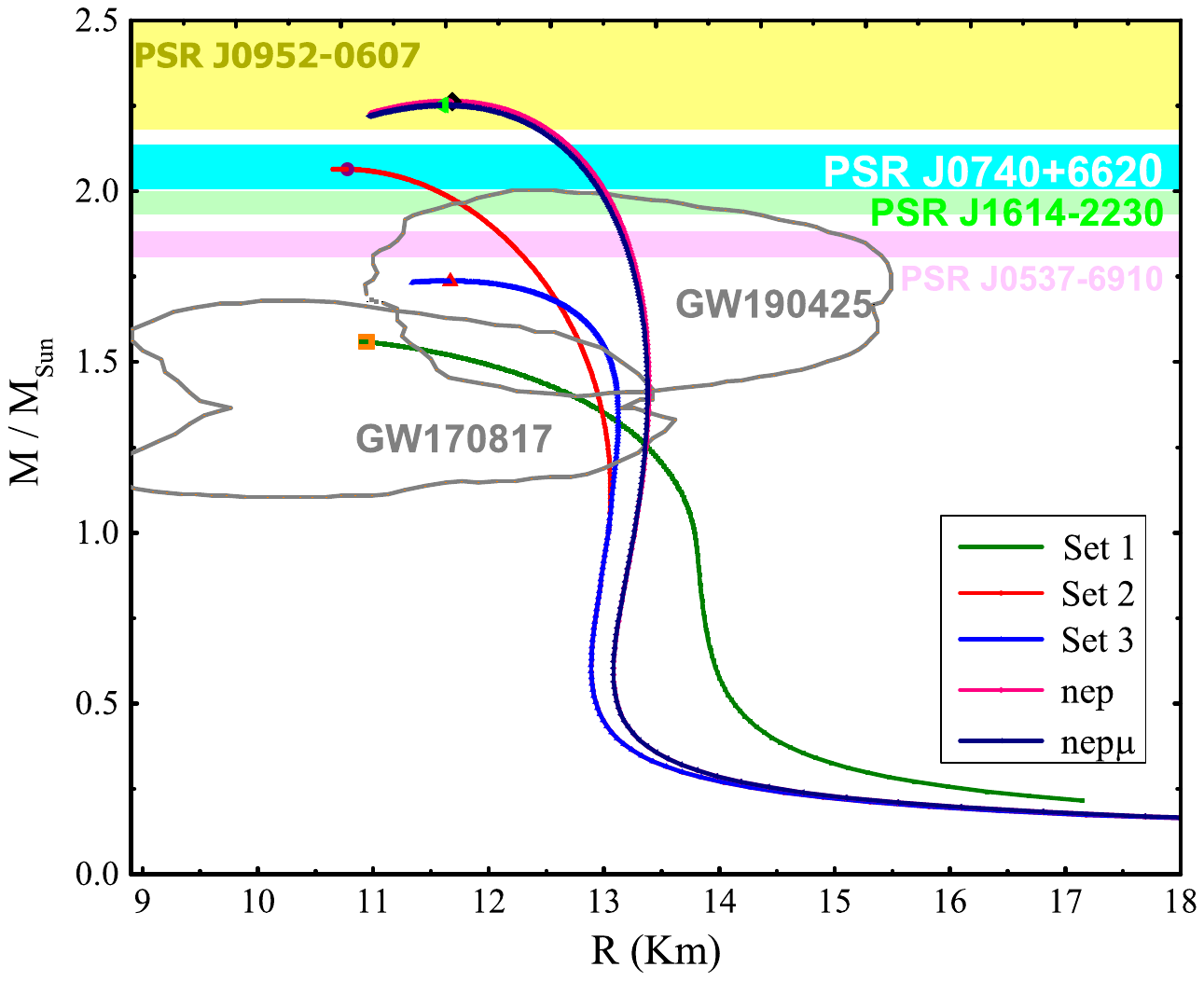}}
	\hspace{1mm}
	\subfigure[ ]{\label{fig:p-nb-tot}
		\includegraphics*[width=0.48\linewidth, height=0.23\paperheight]{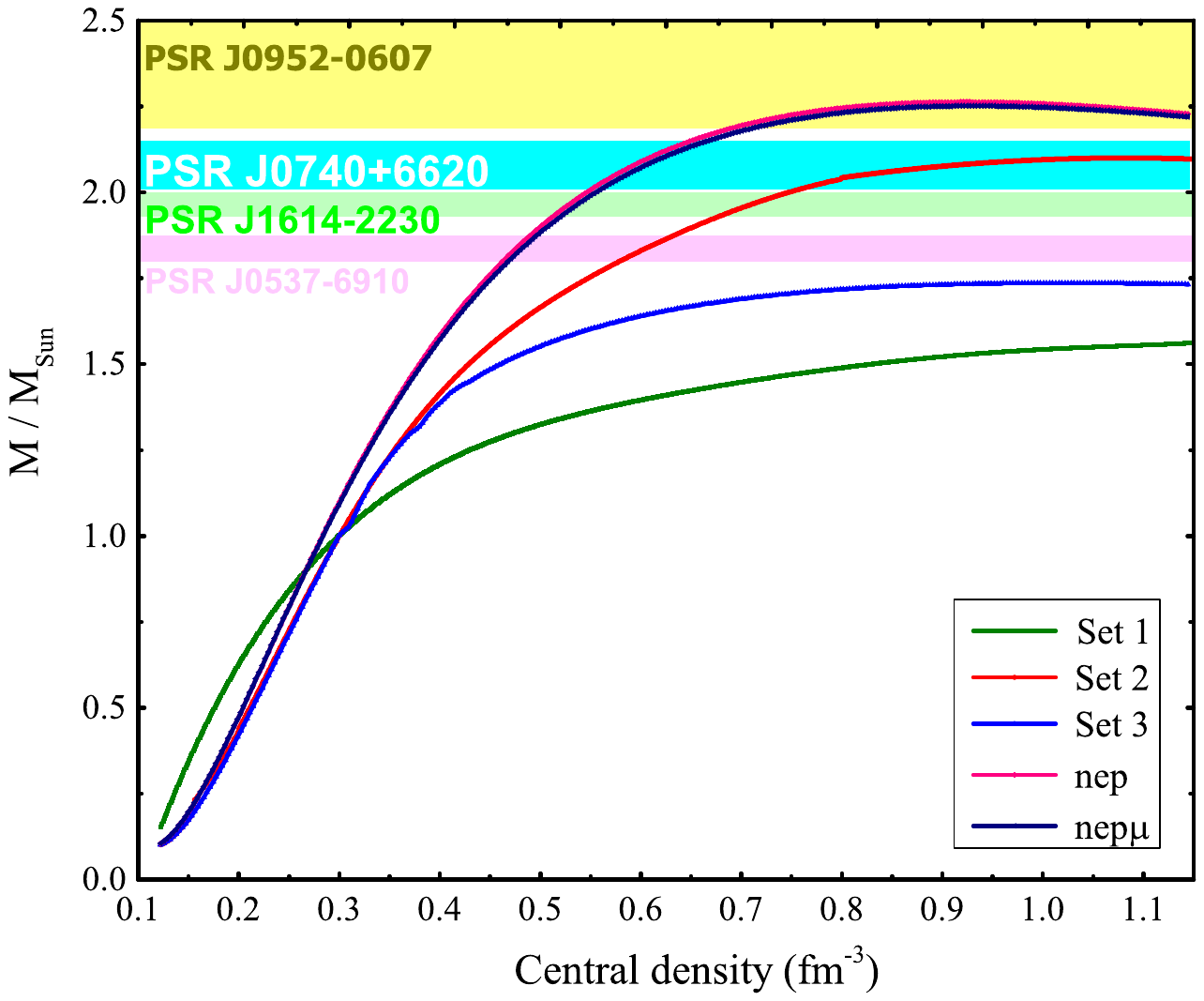}}
	\hfill \\
	\caption{(a): The mass-radius relation (left panel), and (b): gravitational mass as a function of the central baryon density (right panel) in the case of nucleon stars (upper curves) and hyperon stars (lower curves) in compared with observational data (color lines).}
	{\label{figfig}}
\end{figure}

\begin{figure}
	\centering
	\subfigure[]{\label{fig:11}
		\includegraphics*[width=.48\linewidth, height=0.23\paperheight]{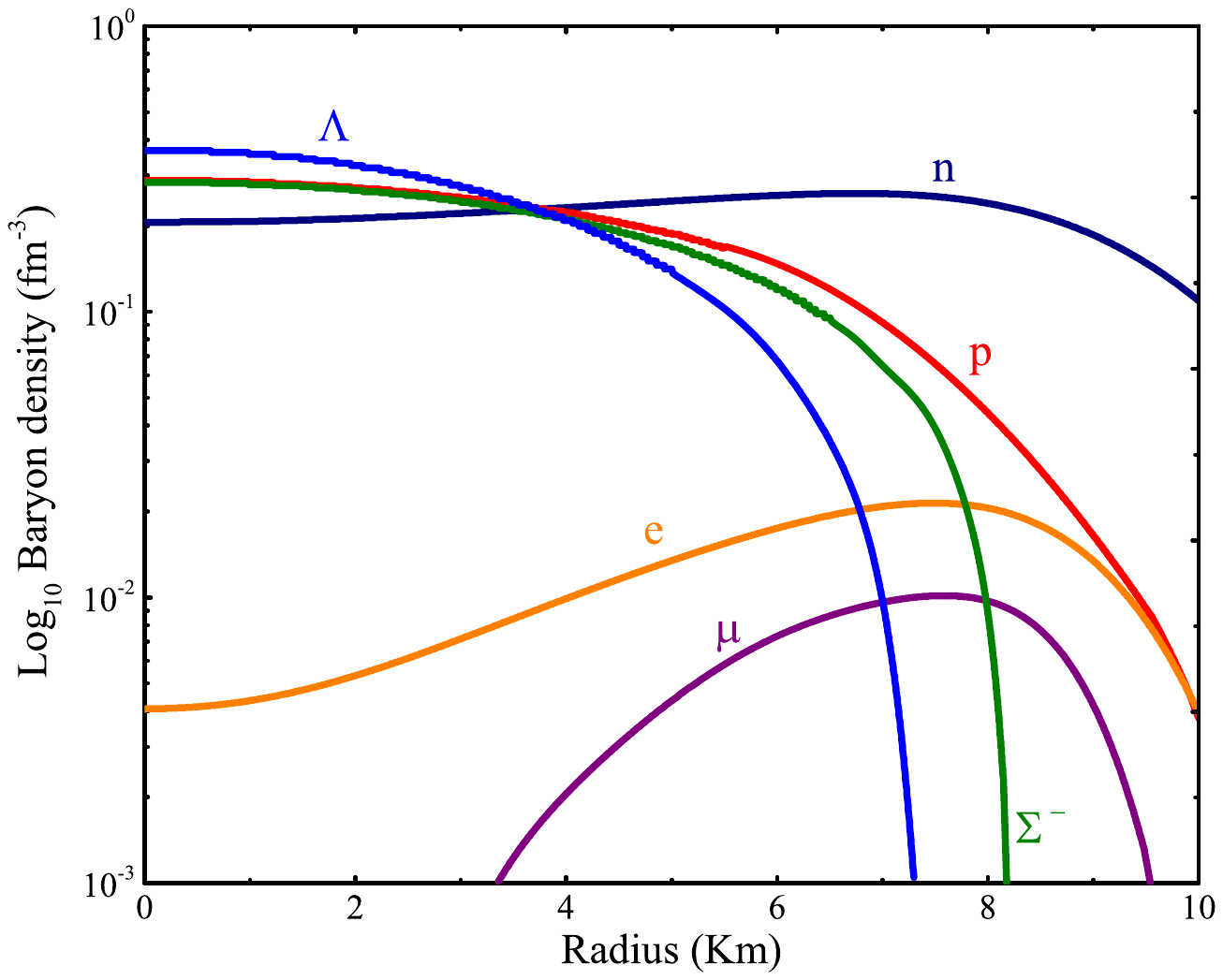}}
	\hspace{1mm}
	\subfigure[ ]{\label{fig:22}
		\includegraphics*[width=.48\linewidth]{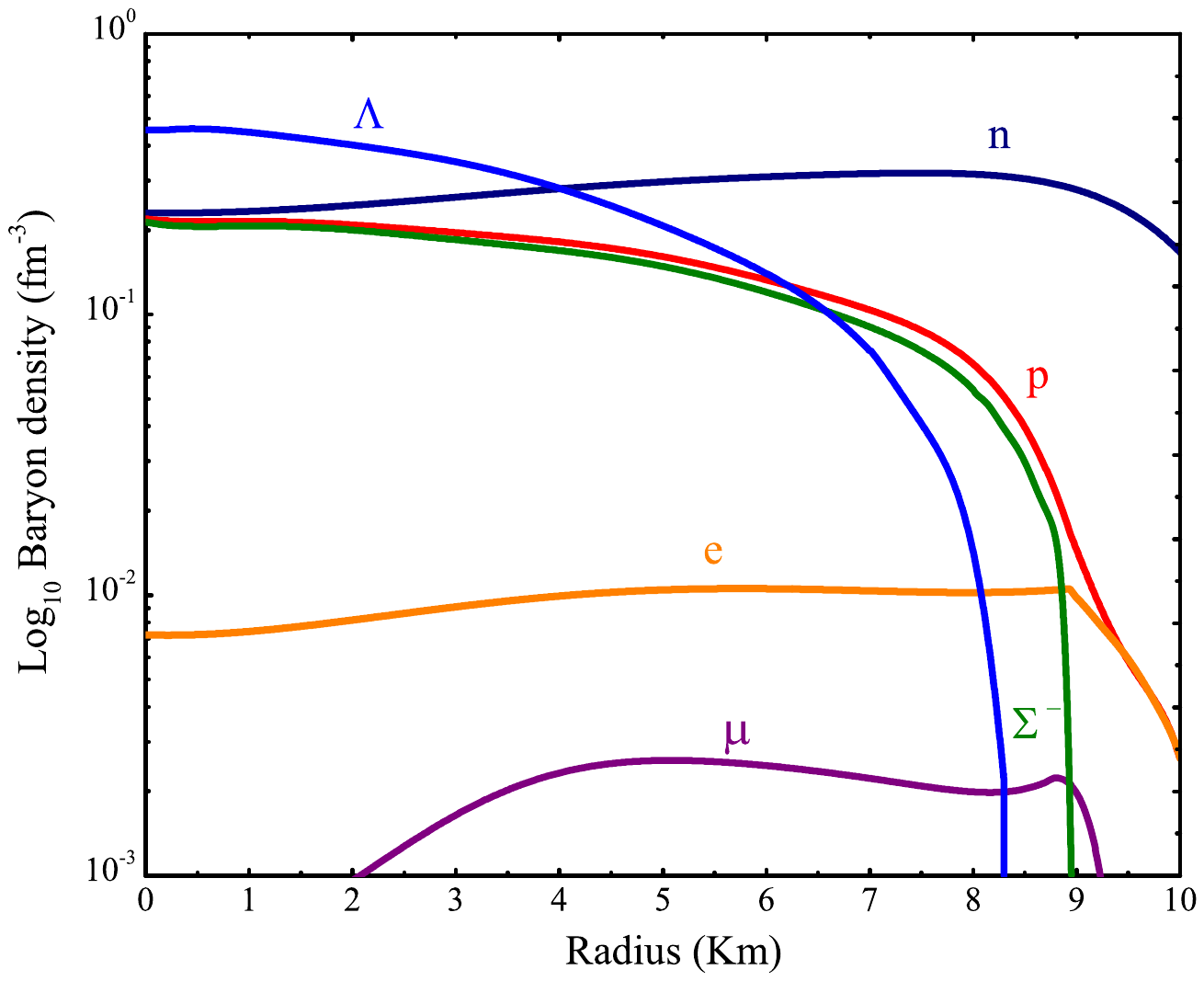}}
	\hspace{1mm}
	\subfigure[ ]{\label{fig:33}
		\includegraphics*[width=.48\linewidth]{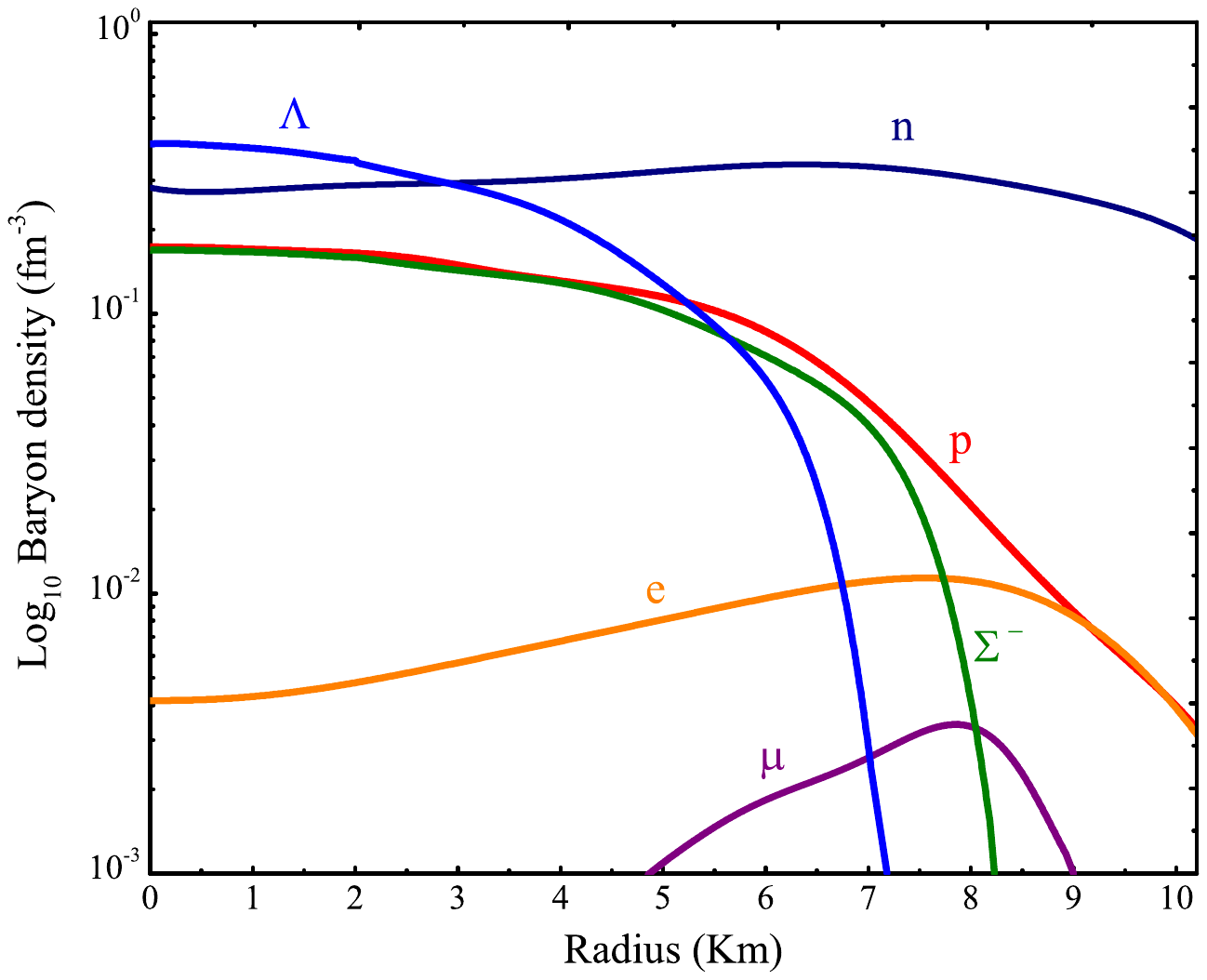}}
	\hspace{1mm}
	\caption{Baryon density as a function of stellar radius for npYe$\mu$ system using coupling constants of (a): set~1, (b): set~2, and (c): set~3.}
	\label{4fig2}
\end{figure}

The Tolman-Oppenheimer-Volkoff (TOV) Equation \cite{Tolman:1939jz, Oppenheimer:1939ne}, are used to determine the stability of an EOS against gravitational collapse. These equations relate the change in pressure with radius to the state variables of the EOS. To solve the equations, the boundary condition that the pressure is zero at the surface of the star is applied. The TOV equations are:
\begin{equation}
	\label{TOV1}
	\dfrac{dP(r)}{dr} = - \dfrac{GM(r) \varepsilon(r)}{c^{2}r^{2}}(1+ \dfrac{P(r)}{\varepsilon(r)})(1+\dfrac{4\pi r^{3} P(r)}{M(r)c^{2}}) (1- \dfrac{2GM(r)}{rc^{2}})^{-1},
\end{equation}
and
\begin{equation}
	\dfrac{dM(r)}{dr} = \dfrac{4\pi\varepsilon(r)r^{2}}{c^{2}}.
\end{equation}
Where $P(r)$, $\varepsilon(r)$, and $M(r)$ represent the pressure, energy density and mass of the star, respectively.
To solve the TOV equations, we require the relation between pressure and energy density. In this work, we use different equations of state (EOS) as discussed in the previous section for the core of the star, and BPS equation of state \cite{PPS} for the crust of the neutron star (NS), to solve the TOV equations.
We assume spherical symmetry and zero pressure gradient $ \dfrac{dP}{dr} $ at the center of the star (r=0). By using the Runge-Kutta method to solve the TOV differential equations with the boundary condition (P=0) at the surface of the star, we can determine the pressure at an arbitrary radius. Subsequently, by utilizing the given equation of state, we can calculate the energy density.
Using calculated $ \varepsilon(r) $, we can then determine the mass within a given radius $R^{'}$ as
\begin{equation}
	M(R^{'}) = \int_0^{R^{'}} 4\pi r^{2} \varepsilon(r)dr.
\end{equation} 
The total mass of the star is given by $ M(R) $, where $R$ is the radius of the star. By varying the central density, $\rho_{C}$, as an initial input, and repeating the integration for different values the mass-radius (M-R) relation is determined. Consequently, the maximum mass and corresponding radius of the neutron star can be found. In Fig. $\ref{figfig}$, we display the M-R relation and gravitational mass as a function of the central baryon density of NS. This include NSs consisting of nep, nep$\mu$ (using coupling constants of set~2) and, nep$\mu\Sigma^{-}\Lambda $ matter of each set. The results are compared with the masses of the heaviest known pulsars, including; PSR~J0740+6620 \cite{Fonseca:2021wxt}, PSR~J0952-0607 \cite{Romani:2022jhd}, PSR~J1614-2230 \cite{Demorest:2010bx}, PSR~J0537-6910 \cite{Ho:2015vza,Ho:2020vxt}, and binary neutron star observed in the GW190425, and GW170817 gravitational events \cite{LIGOScientific:2020aai,LIGOScientific:2018cki, Lim:2019som,Malik:2018zcf}. 
The observational constraints are represented by color coded regions in this figure. The bold points in the figure indicate the maximum mass, and the numerical values of the masses and radii are given in Table $\ref{table:m-r}$. The nep and nep$\mu$ diagrams shows that due to the low muon abundance in NSs, their effect on the M-R relation is negligible. The fraction of muons in NSs is typically less than 10 $\% $, and their contribution to the pressure is small in comparison to nucleons and hyperons. We can observe that QCDSR coupling constant can create a stiff enough EOS compare to set~1, thereby enabling the existence of massive and larger NSs. The outcomes of our sets are in good agreement with the most massive observed objects.
\begin{table}[H]
	\caption{The maximum masses ($M_{max}/M_{\odot}$) and radii ($ R $) values of stellar matter for each set and observed object.}
	\label{table:m-r}
	\small
	\begin{center}
		\begin{tabular}{|| c || c | c | c ||}
			\hline
			Sets &$\rho_{central}$ $(fm^{-3})$ & $M_{max}/M_{\odot}$ & Radius ($Km$)  \\ 
			\hline\hline
			\multicolumn{4}{||c||}{Nucleonic Matter} \\
			\cline{1-4}
			nep & $0.92$ & $2.26$ & $11.65$  \\ 
			nep$\mu$ & $0.93$ & $2.25$ & $11.61$  \\
			\hline\hline
			\multicolumn{4}{||c||}{Hyperon Matter} \\
			\cline{1-4}
			Set 1 & $1.14 $ & $1.56$ & $10.89 $  \\ 
			Set 2 & $1.11$ & $2.07 $ & $10.72$  \\ 
			Set 3 & $1.01$ & $1.74$ & $11.64$  \\ 
			\hline\hline
			\multicolumn{4}{||c||}{Observed Data} \\
			\cline{1-4}
			PSR~J0952-0607 \cite{Romani:2022jhd} & - & $2.35 \pm 0.17 $& -  \\ 
			PSR~J0740+6620 \cite{Fonseca:2021wxt} & - & $2.07 \pm 0.07 $& $12.39^{+1.30}_{-0.98}$  \\ 
			PSR~J1614-2230  \cite{Demorest:2010bx} & -  & $1.97 \pm 0.04 $& -  \\ 
			PSR~J0537-6910 \cite{Ho:2015vza,Ho:2020vxt} & -  & $1.83 \pm 0.04 $& -  \\
			GW190425 \cite{LIGOScientific:2020aai} & -  & $1.40 - 1.90 $& $11.00 - 15.00$   \\
			GW170817 \cite{LIGOScientific:2018cki, Lim:2019som,Malik:2018zcf} & -  & $1.20 - 1.70 $& $8.80 - 13.50$   \\
			\hline
		\end{tabular}
	\end{center}
\end{table}
We have illustrated the distribution of baryons within neutron stars, as depicted in Fig. \ref{4fig2}. Approximately 1.5 to 2.5 km in the surface of stars are predominantly composed of nucleons. Moving further in the presence of muon particles becomes significant.
Closer to the core, the distribution changes. From the center of the stars extending up to about 7 km in stars sigma and lambda baryons are prevalent. Nucleons and hyperons are almost equally distributed within the inner core of these stars. Additionally, small amounts of leptons are observed in these inner core.
This detailed mapping of baryon density provides valuable insights into the complex structure and composition of neutron stars, enhancing our understanding of these fascinating astronomical objects.

\subsection{Calculation of Tidal deformability and Love number}
\begin{figure}
	\centering
	\includegraphics[width=.48\linewidth]{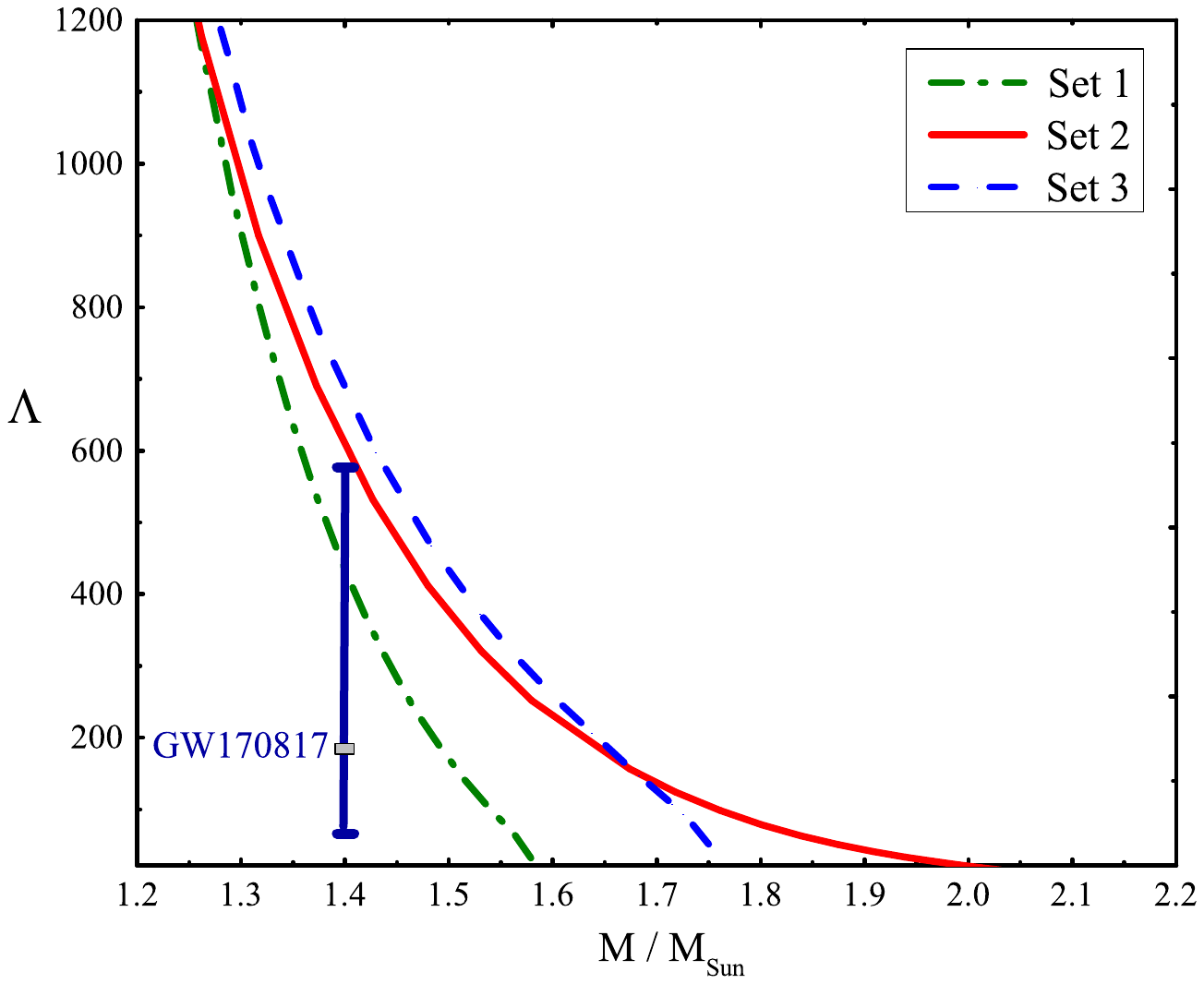}
	\caption{Dimensionless tidal deformability as a function of gravitational mass for NSs at the presence of $\Sigma^{-}$ and $\Lambda$ hyperons for set 1 to set 3.}
	\label{fig:tidal-tot}
\end{figure}

 General relativity predicts that moving bodies distort the surrounding space-time. Oscillations in their mass or motion can produce gravitational waves, which are ripples in space-time. The merging of massive binary systems, such as
 black holes or NSs, is one of the most powerful sources of gravitational waves. The discovery of gravitational waves, which was achieved by the advanced Laser Interferometer Gravitational wave Observatory (LIGO) detector \cite{LIGOScientific:2016aoc}, is a significant milestone in astrophysics/cosmology. This was the first direct observation of these ripples in spacetime, which are predicted by the general theory of relativity. It was caused by the inspiral and coalescence of two black holes. It is expected that LIGO \cite{LIGOScientific:2016aoc}, VIRGO \cite{Virgo:11}, and KAGRA \cite{Kagra:11} will also detect gravitational waves from binary NSs (BNSs). This will provide valuable insights into the properties of highly compressed baryonic matter. Many studies have suggested that the tidal effects of BNSs can be measured by the current generation of gravitational wave (GW) detectors.
 A dimensionless parameter was introduced by the mathematician A. E. H. Love in Newtonian theory \cite{Love:1912}
 to describe the tidal deformation of the Earth caused by the gravitational attraction of the Moon and the Sun. This theory was later extended to general relativity \cite{Damour:2012yf, Binnington:2009bb}, where it was found that there are electric and magnetic types of dimensionless gravitational Love numbers that characterize the tidal fields associated with the gravito-electric and gravito-magnetic interactions. The tidal deformability parameter $\lambda$ is dependent on the EOS through the NS's radius and Love number $\mathcal{K}_{2}$. Flanagan and Hinderer \cite{Flanagan:2007ix} provided the following expression for $\lambda$:
\begin{equation}
	\label{lambda2}
	\lambda = \frac{2}{3}  \mathcal{K}_{2}  R^{5}, 
\end{equation}
the dimensionless tidal deformability $\varLambda$ is related to the compactness parameter $C = \dfrac{M}{R}$ as,
\begin{equation}
	\label{Lambdabig}
	\varLambda = \dfrac{2~\mathcal{K}_{2}}{3~ C^{5}},
\end{equation}
where M and R represent the mass and radius, respectively, of the isolated spherical star. We use units in which c = G = 1.
To calculate the deformability parameter $\lambda$, it is necessary to obtain the Love number $\mathcal{K}_{2}$, which is the key quantity of deformation due to the gravitational attraction between the binary stars. To calculate the Love number, the TOV equations must be solved iteratively along with the following first-order differential equation \cite{Hinderer:2007mb}.
The initial boundary condition is $y(0) = 2$.
\begin{equation}
	r \dfrac{dy(r)}{dr} + y(r)^{2} + y(r) F(r) + r^{2} Q(r) = 0, 
\end{equation}
where,
\begin{equation}
	F(r) = \dfrac{r - 4\pi r^{3} [\varepsilon(r) - P(r)]}{r - 2M(r)},	
\end{equation}
and,
\begin{equation}
	Q(r) = \dfrac{4\pi r \Big(5\varepsilon(r) + 9 P(r) + \dfrac{\varepsilon(r) + P(r)}{\frac{\partial P(r)}{\partial \varepsilon(r)}}  - \frac{6}{4 \pi r^{2}} \Big)  }{r - 2M(r)} - 4 \Big[\dfrac{M(r) + 4\pi r^{3} P(r)}{r^{2}(1- \frac{2M(r)}{r})}\Big]^{2}.
\end{equation}
In $y = y_{2}(R)$, the electric tidal Love numbers are given as below \cite{Damour:2009wj}:
\begin{equation}
	\label{Love}
	\begin{split}
		\mathcal{K}_{2} = & ~ \frac{8}{5} (1-2C)^{2} C^{5} \Big[2C(y_{2} - 1) - y_{2} +2\Big] \Biggl\{2C\Big(4(y_{2} +1)C^{4} + (6y_{2} -4)C^{3} + (26-22y_{2})C^{2} + 3(5y_{2} - 8)C  \\
		& - 3y_{2} + 6\Big) - 3(1-2C)^{2} \Big(2C(y_{2} -1) - y_{2} + 2\Big)  log (\dfrac{1}{1 - 2C})\Biggr\} ^{-1} .\\
	\end{split}
\end{equation}
We used equations (\ref{lambda2}) to (\ref{Love}) to calculate the tidal deformability and Love number of compact stars. Fig. \ref{fig:tidal-tot} shows dimensionless tidal deformability as a function of gravitational mass for NSs at the presence of $\Sigma^{-}$ and $\Lambda$ hyperons for set 1 to set 3. Also, The radius, tidal deformability, Love number and, the compactness values foe the NSs with 1.4  ~$M_{\odot}$ mass are reported in Table \ref{table:1.4M}. As can be seen from table \ref{table:1.4M}, our results are in good agreement with the GW170817, and GW190425 gravitational events data and other papers.

\begin{table}[H]
	\caption{ The properties of the 1.4~$M_{\odot}$ NS at the presence of $\Sigma^{-}$ and $\Lambda$ hyperons for different classes of the coupling constants, with the results compared to those of other references.}
	\label{table:1.4M}
	\small
	\begin{center}
		\begin{tabular}{|| c || c | c | c | c ||}
			\hline
			Sets & Radius ($Km$) & $\varLambda$ & $\mathcal{K}_{2}$ & C \\ 
			\hline\hline
			Set 1 & $12.73$ & $437$ & $0.07$ & $0.16$\\ 
		    Set 2 & $12.32$ & $609$ & $0.12$ & $0.16$\\ 
		    Set 3 & $13.12$ & $685$ & $0.09$ & $0.16$\\ 
			GW170817 & $13_{-1.18}^{+0.72} $ & $190_{-120}^{+390}$ & - & - \\ 
			GW190425 & $ \leq 15 $ & - & - & - \\ 
				Ref. \cite{Khanmohamadi:2020wcf} & $12.42$ & $469.98 $ &$0.09$ & $0.16$\\ 
			Ref. \cite{Kumar:2016dks} & $13.69$ & $ 839.04$ &$0.09$ & $0.15$\\ 
				Ref. \cite{Tong:2024egi} & $13.14$ & $ 597$ &$-$ & $-$\\ 
			Ref. \cite{Sedaghat:2024bnj} & $\leq$ $12.23$  &$\leq$ $887.92$  & - & $\leq$ $0.30$\\ 
			\hline
		\end{tabular}
	\end{center}
\end{table}
\section {CONCLUSION}\label{sec:four}
	
This paper examines the impact of baryon-meson coupling constants on the structure of neutron stars, utilising the relativistic mean-field (RMF) theory as the theoretical framework. Our findings indicate that the energy density and pressure of neutron star matter increase with the QCDSR coupling constants, resulting in a stiffer equation of state and a higher mass comparable to the most massive neutron stars observed in recent years. The utilisation of QCDSR coupling constants modifies the equation of state, thus resolving the hyperon puzzle problem. This enables the observation of a mass exceeding 2 solar masses, even in the presence of hyperons within a neutron star. Furthermore, the numerical results of tidal deformability and Love number values are in accordance with observational data.
 
Finally, it should be noted that the coupling constants employed in this study were derived in a vacuum, yet the results obtained were reasonably satisfactory. It is recommended that future research focus on the acquisition of these coupling constants within matter. It is hypothesized that this will result in slight improvements to the results obtained, although the observational data and the alignment of calculations with the QSDSR coupling constants suggest that even these coupling constants in matter, up to high densities, will not deviate significantly from the reported range. It is anticipated that these coupling constants will be capable of determining the lower limit in vacuum with greater certainty, and at high densities, they will reach the upper values of this range.
	
	\section*{ACKNOWLEDGEMENTS}
K. Azizi thanks Iran national science foundation (INSF) for the partial financial support provided under the elites Grant No. 4037888. He is also grateful to the CERN-TH department for their support and warm hospitality.


\end{document}